\documentclass[aps,prl,twocolumn,showpacs,superscriptaddress,preprintnumbers,amsmath,amssymb]{revtex4}

\usepackage{graphicx}
\usepackage{dcolumn}
\usepackage{bm}
\usepackage{amsmath}
\usepackage{textcomp}
\usepackage[latin1]{inputenc}

\begin{document}
\preprint{APS/123-QED}


\title{Weak localization of light in superdiffusive random systems}

\author{Matteo Burresi}
\email[]{burresi@lens.unifi.it} \affiliation{European Laboratory
for Non-linear Spectroscopy (LENS), 50019 Sesto Fiorentino (FI),
Italy.} \affiliation{Istituto Nazionale di Ottica (CNR-INO), Largo
Fermi 6, 50125 Firenze (FI), Italy.}

\author{Vivekananthan Radhalakshmi}
\affiliation{European Laboratory for Non-linear Spectroscopy
(LENS), 50019 Sesto Fiorentino (FI), Italy.}
\affiliation{Universit\`a di Firenze, Dipartimento di Fisica e
Astronomia, 50019 Sesto Fiorentino (FI), Italy.}

\author{Romolo Savo}
\affiliation{European Laboratory for Non-linear Spectroscopy
(LENS), 50019 Sesto Fiorentino (FI), Italy.}
\affiliation{Universit\`a di Firenze, Dipartimento di Fisica e
Astronomia, 50019 Sesto Fiorentino (FI), Italy.}

\author{Jacopo Bertolotti}
\affiliation{Universit\`a di Firenze, Dipartimento di Fisica e
Astronomia, 50019 Sesto Fiorentino (FI), Italy.}
\affiliation{MESA+ Institute for Nanotechnology, University of
Twente, 7500 AE Enschede, The Netherlands.}

\author{Kevin Vynck}
\email[]{vynck@lens.unifi.it} \affiliation{European Laboratory for
Non-linear Spectroscopy (LENS), 50019 Sesto Fiorentino (FI),
Italy.} \affiliation{Universit\`a di Firenze, Dipartimento di
Fisica e Astronomia, 50019 Sesto Fiorentino (FI), Italy.}

\author{Diederik S. Wiersma}
 \affiliation{European Laboratory
for Non-linear Spectroscopy (LENS), 50019 Sesto Fiorentino (FI),
Italy.} \affiliation{Istituto Nazionale di Ottica (CNR-INO), Largo
Fermi 6, 50125 Firenze (FI), Italy.}

\date{\today}

\begin{abstract}
L\'{e}vy flights constitute a broad class of random walks that
occur in many fields of research, from animal foraging in biology,
to economy to geophysics. The recent advent of L\'{e}vy glasses
allows to study L\'{e}vy flights in controlled way using light
waves. This raises several questions about the influence of
superdiffusion on optical interference effects like weak and
strong localization. Super diffusive structures have the
extraordinary property that all points are connected via direct
jumps, meaning that finite-size effects become an essential part
of the physical problem. Here we report on the experimental
observation of weak localization in L\'{e}vy glasses and compare
results with recently developed optical transport theory in the
superdiffusive regime. Experimental results are in good agreement
with theory and allow to unveil how light propagates inside a
finite-size superdiffusive system.
\end{abstract}

\maketitle L\'{e}vy flights are maybe the most general class of
random walks, of which the commonly known Brownian motion is a
limiting case \cite{mandelbrot_book,Klafternature}. They are
governed by L\'{e}vy statistics\cite{Levy}, which have the
fascinating property that, depending on the value of one control
parameter, they can exhibit a diverging variance \cite{nolan}.
This leads to a phenomenon called superdiffusion, being a
diffusive process in which the mean square displacement increases
faster than linear in time
\cite{Drysdale1998,metzler_random_2000}. L\'{e}vy flights are
common in nature and appear, for instance, in animal food searches
\cite{Viswanathan1999,Bartumeus2005}, laser cooling of cold atoms
\cite{CohenTann1994}, evolution of the stock market
\cite{Mandelbrot1963}, astronomy \cite{boldyrev}, random lasers
\cite{kumar1} and turbulent flow \cite{Swinney1993}.

The recent development of L\'{e}vy glasses
\cite{barthelemy_levy_2008} and carefully prepared hot atomic
vapours \cite{mercadier_levy_2009}, have allowed the observation
of L\'{e}vy flights of light waves and the resulting
superdiffusion process. Since interference effects play a dominant
role in light transport, this raises the natural question how
interference influences optical superdiffusion - a concept which
has not been addressed so far. In regular - diffusive - disordered
optical systems, interference leads to speckle correlations, and
weak and strong localization effects, which all have been studied
extensively over the last two decades
\cite{sheng,akkermans_mesoscopic_2007}. Localization, in
particular, leads to a complete halt of transport that can confine
light waves in random patterns. Since the superdiffusion induced
by L\'{e}vy statistics tends to enhance transport, one would
expect it to counter-act localization induced confinement.

Among all interference phenomena in random optical materials,
maybe the most robust is that of weak localization \cite{sheng}.
It is observed in the form of a cone of enhanced backscattering,
which contains information on the path length distribution deep
inside the random system, and which has been observed in recent
years from several diffusive random structures
\cite{Kuga_1984,albada_observation_1985,PhysRevLett.55.2696,PhysRevLett.75.1739,muskens_broadband_2011,PhysRevLett.83.5266,PhysRevLett.92.033903,PhysRevLett.94.183901}
Some of us have recently shown theoretically that weak
localization, or coherent backscattering, can be observed from
L\'{e}vy glasses and that a superdiffusive approximation can be
used to predict its behaviour \cite{bertolotti_multiple_2010}. In
this paper we report on the experimental observation of weak
localization from superdiffusive materials, which constitutes the
first observation of an interference effect in transport based on
L\'{e}vy statistics. We find a good agreement with superdiffusive
transport theory and show how the backscattering cone can be used
to extract the Green's function in a L\'{e}vy glass. Contrary to
regular diffusive media, in a L\'{e}vy glass, light from an {\em
arbitrary} depth inside the medium has a nonvanishing probability
to couple directly to the surrounding environment. This latter
property makes light scattering from L\'{e}vy glasses complex, and
has important consequences for its (back)scattering properties.

\begin{figure}[t]
\includegraphics[width=8.5cm]{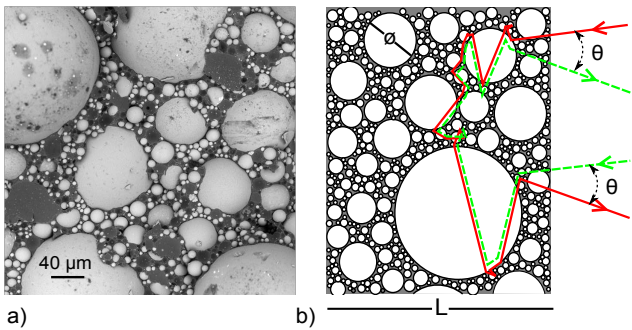}
\caption{ (a) Electron micrograph of the interior of a L\'{e}vy
glass. 
(b) Sketch representing the optical mechanism laying behind the
coherent backscattering cone in a Levy glass. $L$=300
\hbox{\textmu}m is the thickness of the sample and $\o$ the
diameter of the spheres.}\label{fig1}
\end{figure}
\begin{figure*}[t!]
\includegraphics[width=17.5cm]{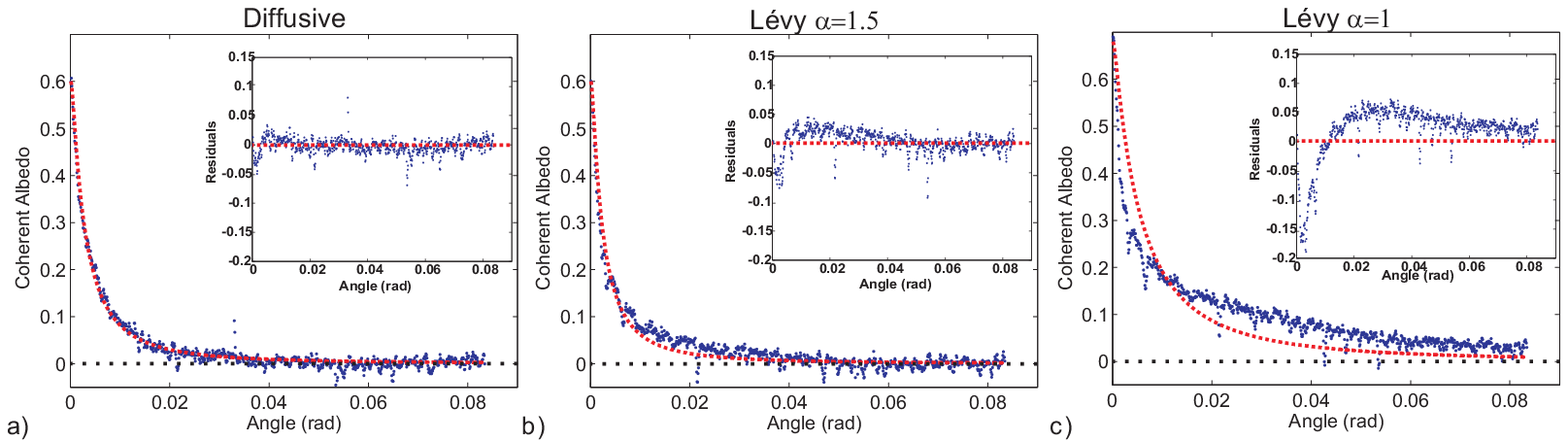}
\caption{ (a), (b) and (c) Measured coherent backscattering cone
from diffusive samples and L\'{e}vy glasses with $\alpha=1.5$ and
$\alpha=1$, respectively. In red the fit to the experimental data
according to the standard diffusion theory. In the insets,
residuals of the fits.}\label{fig2}
\end{figure*}

The L\'{e}vy glasses under investigation are made of jammed-packed
microscopic glass spheres, whose diameter $(\o)$ varies almost
over two order of magnitudes (from 5 to 230 \hbox{\textmu}m)
following a power-law distribution $p(\o)\propto \o^{-(\beta+1)}$,
with $\beta$ adjustable parameter. These spheres are embedded in a
polymeric matrix which matches their refractive index
($\mathrm{n=1.52}$) and in which $\mathrm{Ti O_2}$ nanoparticles
(average diameter 280 nm) have been dispersed (see Fig. 1(a))
\cite{bertolotti_engineering_2010}. Because of their refractive
index ($\mathrm{n=2.4}$) higher than the polymer, these
nanoparticles act as point scatterers which are not homogenously
distributed throughout the sample due to the presence of the glass
spheres (see Fig. 1(b)). As a result, light transport is dominated
by the long "jumps" that light performs propagating through the
microscopic spheres. The step length distribution that light
performs in L\'evy glasses follows a power-law decay as
$p(l)\propto l^{-(\alpha+1)}$, where $\alpha$ is related to
$\beta$ as $\alpha=\beta-1$ for an exponential sampling of the
diameter distribution \cite{bertolotti_engineering_2010}. 
By controlling the diameter distribution of the spheres in a
L\'{e}vy glass we can control $\alpha$ and, thus, the degree of
superdiffusion  of the material. For $\alpha\geq 2$ the system is
diffusive, whereas for $0<\alpha< 2$ the system is superdiffusive.
However, in real systems the finite size of the sample truncates
the step-length distribution to the largest sphere diameter,
enriching the physics behind the transport properties
\cite{barthelemy_role_2010,
groth_transmission_2011,buonsante_transport_2011}. 
In previous publications
\cite{barthelemy_levy_2008,bertolotti_engineering_2010} L\'{e}vy
glasses were fabricated between two microscope slides. In
contrast, in this work we remove one of the two slides to reduce
undesirable reflections which affect the quality of the
measurements. Moreover, the thickness of the sample is
approximately $70$ \hbox{\textmu}m more than the largest sphere,
which means that the step length distribution is truncated on a
length scale that is slightly below the sample thickness.

The setup employed follows a standard scheme for coherent
backscattering experiments. Light
emitted by a HeNe laser (@632 nm) is expanded to collimated beam of
1 cm in diameter to ensure a high angular resolution of the system.
A beamsplitter is
used to separate the backscattered light from light impinging on
the sample. Subsequently, the backscattering cone is imaged on a
CCD camera and the use of a polarizer ensures that we observe only
the polarization conserving channel
\cite{akkermans_mesoscopic_2007}. The angular resolution of the
setup has been optimized to properly investigate the different shapes
of the measured cones. During the data acquisition the sample is
nutated to average over different realizations of disorder.

In order to verify the quality of the setup we fabricate a set of
diffusive samples made of $\mathrm{TiO_2}$ and polymer (without
glass spheres) and measured its backscattering cone. The amount of
polymer was adjusted to have the same average (over the total
volume) density of scatterers in both diffusive and superdiffusive
samples. A cross-cut of the measured diffusive cones is shown in
Fig. 2(a) together with a fit (red curve) obtained from the
diffusion theory \cite{PhysRevLett.75.1739}. The retrieved
transport mean free path is $\ell^\star\simeq 19$ \hbox{\textmu}m.
In the inset the residuals are displayed as blue dots. The good
match between experimental data and fit shows the quality of the
experimental setup.

We measured the backscattering cone on two different sets of
superdiffusive samples characterized by $\alpha=1.5$ and
$\alpha=1$. The experimental results are shown in Fig. 2(b) and
(c). By comparing these figures one can notice the rising of the
tail of the cone as the degree of superdiffusion increases, i.e.
when $\alpha$ decreases, following a trend which has been
theoretically predicted in Ref. \cite{bertolotti_multiple_2010}.
To ensure that the L\'{e}vy cones are not merely \emph{broader}
diffusive cones we made a best fit using diffusion theory. The
result is shown in Fig. 2(b) and (c) as red curves and the
residuals of such fits are shown in the inset of Fig. 2(a). It is
clear that the diffusion theory cannot properly describe the
optical properties of L\'{e}vy glasses.

The coherent backscattering cone for the superdiffusive samples
under investigation can be calculated by taking advantage of the
fractional derivative approach developed in
Ref.~\cite{bertolotti_multiple_2010}. The transport of light in a
finite, translationally invariant, superdiffusive system for a
point source at $x_0$ is described by the stationary fractional
diffusion equation~\cite{bertolotti_multiple_2010}:
\begin{equation}
 D_{\alpha} \left( \nabla^{\alpha}_x - k_{\perp}^{\alpha} \right) f \left( x, x_0, \mathbf{k}_{\perp} \right) = - \delta(x-x_0),
\label{eq1}
\end{equation}
where $\nabla^{\alpha}$ is the symmetric Riesz fractional
derivative with respect to spatial derivatives,
$\mathbf{k}_{\perp}$ is the in-plane component of the wavevector
in free space, $f\left( x, x_0, \mathbf{k}_{\perp}\right)$ is the
intensity propagator and $D_\alpha$ is a generalized diffusion
constant. The spatial nonlocality of $\nabla^\alpha$ makes the
definition of boundary conditions non
trivial~\cite{chechkin_first_2003}. In order to model physical
systems, such as L\'{e}vy glasses, the fractional Laplacian
operator can be represented by an $M \times M$ matrix, whose
eigenvalues $\lambda_i$, rescaled as $\lambda_i \rightarrow
\lambda_i (M/L)^\alpha$ with $L$ the slab thickness, and
eigenvectors $\psi_i$ converge to those of the continuum operator
as $M$ goes to infinity~\cite{zoia_fractional_2007}. Absorbing
boundary conditions can then be implemented by reducing the
infinite size matrix to a finite size matrix and Eq. \ref{eq1} be
solved by eigenfunction expansion. The knowledge of the intensity
propagator of the system then makes it possible to calculate
interference effects in a ``superdiffusion'' approximation. In
particular, considering a planewave at normal incidence on the
slab interface and in the Fraunhofer regime, the coherent
component of the albedo is given by the following expression:
\begin{equation}
A_c(\theta) \propto - \sum_{x_1, x_2} F(x_1,x_2,\theta)
\sum_{i=1}^{M} \frac{\psi_i (x_1) \psi_i (x_2)}{\left( \lambda_i -
k_\perp ^{\alpha} \right)},\label{eq2}
\end{equation}
where $F(x_1,x_2,\theta) = P(x_1) P(x_2) P(x_1/\cos{\theta})
P(x_2/\cos{\theta})$ describes the attenuation for the amplitude
of the incident and emergent planewaves in the scattering medium,
$\theta$ is the angle between the incident and emergent planewaves
(see Fig. 1(b)) and $k_\perp=|\mathbf{k}_{\perp}|\simeq
(2\pi/\lambda)\;\theta$, at small angle $\theta$. The amplitude
attenuation $P(l)$ was modelled as a Pareto-like distribution
$P(l)=1$ for $0 \leq l \leq l_c$ and $P(l)=(l_c/l)^{(\alpha+1)/2}$
for $l\geq l_c$, where $l_c$ is the cut-off length, as to closely
follow the step length distribution of real L\'evy glasses
\cite{barthelemy_role_2010}. Internal reflections were neglected.

The results are shown in Fig. 3 for $\alpha=1.5$ and $\alpha=1$,
where the \emph{only} adjustable parameter used is $l_c$. The
inset shows the residuals between theory and experiment, which are
greatly reduced with respect to the diffusive fit (Fig. 2(b) and
(c)). Due to the very long tail of the cones for L\'{e}vy glasses
and the lack of an analytical expression for it, the experimental
incoherent background was set to the one obtained
semi-analytically. It must be pointed out that for the
calculations we employed a step-length distribution which was not
truncated, in contrast with the real system. This is due to the
fact that the propagator $f\left(x,x_0,\mathbf{k}_{\perp}\right)$
of the system has been calculated by considering the sample sa
translational invariant in the in-plane direction and finite in
the longitudinal direction. 
The very good agreement between
experiment and calculation shows that the fractional diffusion
approach can properly describe light interference effects due to
multiple scattering in the superdiffusion approximation.
\begin{figure}[t]
\includegraphics[width=6cm]{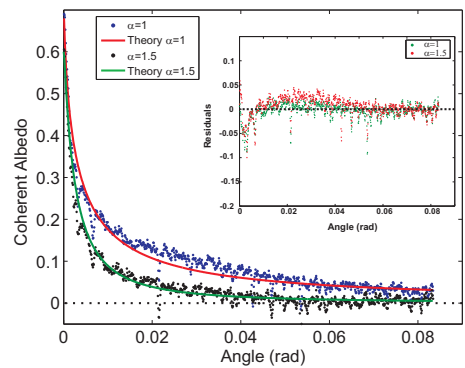}
\caption{Comparison between the calculated superdiffusive cones
obtained with the fractional derivative approach and the measured
L\'{e}vy cones.}\label{fig3}
\end{figure}

The gentle slope of the cone tails arising from a L\'{e}vy glass
is a hallmark of anomalous transport of light. This is a striking
effect since generally the width of the cone is expected to be
inversely proportional to the average step length $\ell$, whereas
in a L\'{e}vy glass the average step length $\ell_{\alpha}$
increases when $\alpha$ decreases. The knowledge of the peculiar
Green's function inside this medium, which dictates the shape of
the cone, might shed light on this phenomenon. The behavior of
such a propagator as a function of the degree of superdiffusion
can be retrieved by a Fourier analysis of the experimentally
observed coherent backscattering. In diffusive media the coherent
albedo can be approximated with the propagator of the system in
reciprocal space. The propagator is taken for a point source at a
distance of a mean free path away from the boundary ($A_c\propto
f(x=\ell,x_0=\ell,k_\perp)$) \cite{akkermans_coherent_1986}. We
expect a similar Fourier analysis on the coherent albedo from
L\'{e}vy glasses to provide information on the intensity
distribution inside a superdiffusive media.
\begin{figure*}[t!]
\includegraphics[width=17.5cm]{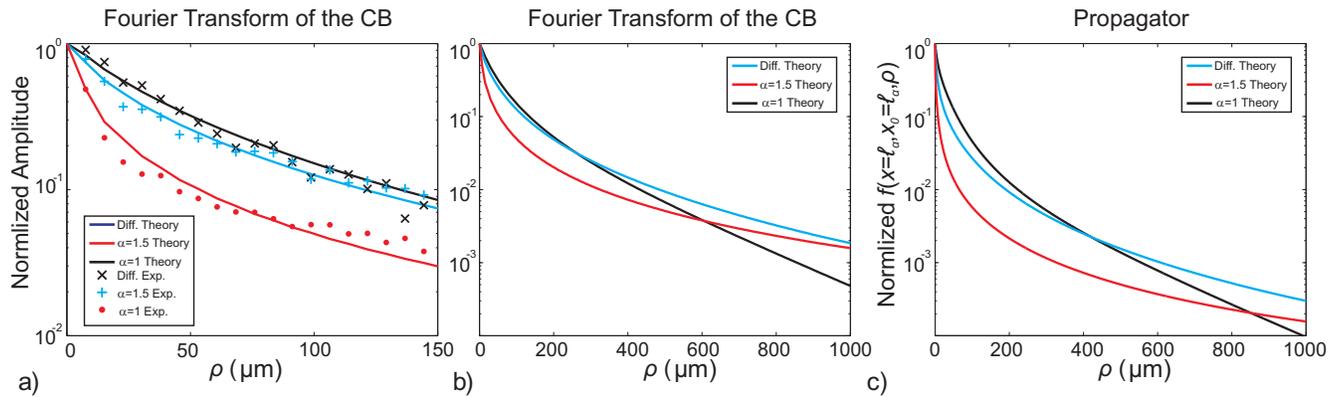}
\caption{(a) Amplitude of the Fourier transform of the measured
and calculated coherent backscattering (CB) for small $\rho$. (b)
and (c) Amplitude of the Fourier transform of the calculated cone
and calculated intensity distribution
$f(x=\ell_\alpha,x_0=\ell_\alpha,\rho)$ for large
$\rho$.}\label{fig4}
\end{figure*}
In Fig. 4(a) the normalized Fourier transform of the measured and
calculated coherent albedo as a function of the in-plane
displacement $\rho=|(\mathbf{r}_i-\mathbf{r}_e)_{\perp}|$ are
shown. These curves are found to be in good agreement, in
particular at relatively small values of $\rho$, and show a
remarkable reshaping as a function of $\alpha$. In Fig 4(b) and
(c) the normalized Fourier transform of the theoretical cone and
the normalized intensity distribution
$f(x=\ell_\alpha,x_0=\ell_\alpha,\rho)$ calculated from the
fractional diffusion approach, respectively, are shown for long
$\rho$. The qualitative agreement between these two figures is
evident, in particular in their dependency on the degree of
superdiffusion. The spatial distribution of the propagator is
dictated by the power-law step length distribution in L\'{e}vy
glasses, which allows light to couple directly to the surrounding
environment from \emph{any} depth inside the sample. This spatial
non-locality applied to an open system such as a L\'{e}vy glass
results in a strong modification of the shape of the propagator as
a function of $\alpha$ \cite{bertolotti_multiple_2010}. Firstly,
the more cusped shape of the
$f(x=\ell_\alpha,x_0=\ell_\alpha,\rho)$ leads to a smaller
in-plane displacement and thus to higher tails in the coherent
albedo. Secondly, the gentle tail of the propagator, which induces
a cusped cone \cite{bertolotti_multiple_2010}, is a consequence of
the fact that light can escape more easily from its local
environment towards large distances.

In conclusion, we have reported on the experimental observation of
weak localization from L\'{e}vy glasses and found strong
deviations from the prediction of the diffusion theory. The
recently developed semi-analytical model for superdiffusive system
is able to reproduce the experimental results with a high degree
of accuracy. From the study of the coherent backscattering cone we
can retrieve the behavior of the Green's function inside real
systems in which the long steps probability is nonvanishing, not
only in L\'{e}vy glasses but possibly also in foams, clouds or
strongly heterogeneous porous media. Further coherent
backscattering studies oriented to investigate the influence of
real systems parameters (e.g., finite size, truncation, quenching)
on the superdiffusive backscattering and their fingerprints in the
dynamics of the backscattering cone can be performed to unravel
the optical properties of real superdiffusive systems.

\begin{acknowledgements}
This work is supported by the European Network of Excellence
Nanophotonics for Energy Efficiency and ENI S.p.A. Novara. This
work is also financially supported by the IIT-SEED Project
Microswim and by the Italian FIRB-MIUR "Futuro in Ricerca" project
RBFR08UH60.
\end{acknowledgements}

\bibliographystyle{prova}

\end{document}